%% file: main.tex
\pdfoutput=1

\documentclass[a4paper]{llncs}

\usepackage{amssymb}
\usepackage{amsmath}
\usepackage[usenames,dvipsnames]{xcolor}
\usepackage[hidelinks]{hyperref}
\usepackage{proof-dashed}
\usepackage{graphicx}
\usepackage{listings}

\input{macros}

\begin{document}
\pagestyle{plain}

\title{Kickstarting Choreographic Programming}
% \subtitle{Models, Foundations, and Implementation}

\author{Fabrizio Montesi}
\institute{University of Southern Denmark\\
\email{fmontesi@imada.sdu.dk}}

\maketitle

\begin{abstract}\input{abstract}\end{abstract}

\section{Introduction}\label{sec:intro}
\input{introduction}

\section{Language models}
\label{sec:models}

\input{models}

\section{Implementation}
\label{sec:implementation}
\input{implementation}

\section{Related Work}
\input{related}

\section{Conclusions and Future Work}
\input{conclusions}

\bigskip

\noindent\textbf{Acknowledgements.}
The author was supported by the Danish Council for Independent Research project \emph{Choreographies for Reliable and 
efficient Communication software} (CRC),
grant no. DFF--4005-00304, and by the EU COST Action IC1201 \emph{Behavioural Types for Reliable Large-Scale 
Software Systems} (BETTY).

\bibliographystyle{abbrv}
\bibliography{biblio}

\end{document}

%% file: macros.tex
\newcommand{\lstcode}[1]{\lstinline~#1~}

\newcommand{\m}[1]{\mathsf{#1}}

\newcommand{\code}[1]{\texttt{#1}}
\newcommand{\lgbox}[1]{\fcolorbox{lightgray}{white}{#1}}

\newcommand{\lto}[1]{\mathrel{\stackrel{{\;\;#1\;\;}}{\mbox{\rightarrowfill}}}}

\definecolor{light-gray}{gray}{0.928}
\newcommand{\pid}[1]{\m{#1}}

\newcommand{\role}[1]{\mathtt{#1}}

\newcommand{\com}[3]{#1\;\code{-\hspace{-0.3mm}>}\;#2:#3}

\newcommand{\swapC}{\simeq_{\mathcal C}}

\definecolor{color:keyword}{rgb}{0.53,0.05,0.05}
\definecolor{color:comment}{rgb}{0.25,0.37,0.75}
\definecolor{color:string}{rgb}{0.87,0.0,0.0}
\usepackage{bold-extra}

\lstdefinelanguage{Jolie}{
morekeywords={
	OneWay,RequestResponse,new,
	main,define,inputPort,outputPort,init,execution,include,
	cset,if,else,csets,interface,type,throws,global,constants,for,
foreach,while,int,double,raw,void,undefined,string,long,bool,any,single,
sequential,concurrent,Jolie,Java,JavaScript,embedded,Location,Protocol,
Interfaces,Aggregates,scope,install,cH,comp,throw,this,default,synchronized,
nullProcess,false,true
},
sensitive=true,
morecomment=[l]{//},
morecomment=[s]{/*}{*/},
morestring=[b]",
otherkeywords={;,|,:}
}

\lstdefinelanguage{Chor}{
morekeywords={
	main,program,protocol,define,public,ask,local,if,else,start,show,
	string,long,bool,int,void,end
},
sensitive=true,
morecomment=[l]{//},
morecomment=[s]{/*}{*/},
morestring=[b]",
otherkeywords={@,;,->,|,:}
}

\newcommand{\setChor}{\lstset{language=Chor}}

%% file: abstract.tex
% We report on some recent efforts by the author and collaborators aimed at the development of \emph{Choreographic 
% Programming}, a programming paradigm for concurrent systems where programs are choreographies of communications.
% 
% Choreographies of communication behaviour have been successfully
% used as powerful tools for the design of concurrent systems.
We present an overview of some recent efforts aimed at the development of 
\emph{Choreographic Programming}, a programming paradigm for the production of concurrent software that is guaranteed 
to be correct by construction from global descriptions of communication behaviour.
% using the same idea for the development of a fully-fledged programming paradigm, called
% \emph{Choreographic Programming}, where programs are guaranteed to have 
% The hallmark characteristic of choreographic programming is that distributed programs, called choreographies,
% are given in terms of global descriptions 
% % We report on the main aspects of formal models, foundations, and a prototype
% % implementation of the paradigm

%% file: introduction.tex
Programming communications among the endpoints in a concurrent system is challenging, because
it is notoriously difficult to predict how nontrivial programs executed simultaneously may interact~\cite{LPSZ08}.
% Leaving programmers to handle this task manually can easily lead to bugs or unnecessary performance bottlenecks.
To mitigate this issue, \emph{choreographies} can be used to give precise specifications of communication behaviour~\cite{wscdl,BPMN}.

A choreography specifies the expected communications among endpoints from a global viewpoint, in
contrast with the standard methodology of giving a separate specification for each endpoint that defines its Input/Output (I/O) actions.
As an example, consider the following choreographic specification
(whose syntax is derived from the ``Alice and Bob'' notation from~\cite{NS78}):
\[
%\begin{array}{l@{\ }l}
\com{\pid{Alice}}{\pid{Bob}}{book};\quad
\com{\pid{Bob}}{\pid{Alice}}{money}
%\end{array}
\]
The choreography above describes the behaviour of two endpoints, $\pid{Alice}$ 
and $\pid{Bob}$. First, $\pid{Alice}$ sends to $\pid{Bob}$ a
$book$; then, $\pid{Bob}$ replies to $\pid{Alice}$ with some $money$ 
as payment.
The motivation for using a choreography as specification is that it is always ``correct by design'', since it 
explicitly describes the intended communications in a system. In other words, a choreography can be seen as a
formalisation of the communication flow intended by the system designer.
% Moreover, 
% each communication is treated as atomic: the sending and receiving 
% actions of the respective sender and receiver endpoints cannot be seen 
% separately.

%Previous work formally investigated how to compile
A choreography can be compiled to the local specifications of the I/O actions that each endpoint should perform~\cite{QZCY07,LGMZ08,CHY12}, as depicted below:
%The idea is to use a choreography to define the behaviour of a system, and then to compile it to 
%(abstract) executable code for each endpoint with an automatic procedure called Endpoint Projection (EPP), with the 
%guarantee that the generated code complies with the originating choreography.
% The authors prove that their respective EPP procedures are formally correct, meaning that the generated endpoint 
% programs implement the communications described in the originating choreography.
% As choreographies are correct by design, it follows that the generated code is also correct by construction.
%We depict this methodology in the following:
\begin{center}
    \begin{tabular}{cccccccccccccccc}
      \begin{tabular}{l}
        \lgbox{Choreography Spec.}
      \end{tabular}
      &
      \quad
      $\lto{
       \begin{array}{c}
        \textit{EPP}
       \end{array}
      }$
      \quad
      &
      \begin{tabular}{c}
        \lgbox{Endpoint Spec.}
      \end{tabular}
      \\[2mm]
      \emph{(correct by design)}
      &
      &
      \emph{(correct by construction)}
    \end{tabular}
\end{center}
% 
% The major benefit of the methodology above is that the generated endpoint code is ``correct by construction'', since a
% 
In the methodology above, an Endpoint Projection (EPP) procedure is used to generate the specifications for each 
endpoint starting from a global choreographic specification.
The endpoint specifications are therefore correct by construction, because they are computed from a correct-by-design choreography.
%Since endpoint specifications are generated from a correct-by-design choreography, endpoint specifications are also correct by construction
%Endpoint specifications are ``correct by construction'', based on the fact that their originating choreography is assumed to be a formalisation of the 
A major consequent benefit is that such endpoint specifications are also deadlock-free, because I/O actions cannot 
be defined separately in choreographies and are therefore always paired correctly in the result of EPP.
% In general, since choreographies are explicit representations of the desired communication behaviour, we say that the 
% generated code is correct by construction.
% 

In this paper, we give an overview of some recent results by the author and collaborators aimed at applying the 
choreography-based methodology as a fully-fledged programming paradigm, rather
than as a specification method.
In this paradigm, called \emph{Choreographic Programming}, choreographies are concrete programs and EPP is a compiler 
targeting executable distributed code:
\begin{center}
    \begin{tabular}{cccccccccccccccc}
      \begin{tabular}{l}
        \lgbox{Choreography Program}
      \end{tabular}
      &
      \quad
      $\lto{
       \begin{array}{c}
        \textit{EPP (compiler)}
       \end{array}
      }$
      \quad
      &
      \begin{tabular}{c}
        \lgbox{\begin{tabular}{c}Executable\\ Endpoint Programs\end{tabular}}
      \end{tabular}
      \\[4mm]
      \emph{(correct by design)}
      &
      &
      \emph{(correct by construction)}
    \end{tabular}
\end{center}
Ideally, this methodology will allow developers to program systems from a global 
viewpoint, which is less error-prone than writing endpoint programs directly, and then to obtain executable code 
that is correct by construction.

To kickstart the development of choreographic programming, we are 
interested in finding suitable language models (\S~\ref{sec:models}) and their implementation (\S~\ref{sec:implementation}). We discuss them in the remainder of the paper, following the syntax from~\cite{M13:phd}.

%% file: models.tex
In~\cite{CM13} we present the Choreography Calculus (CC), a language model for choreographic programming that follows
the correct-by-construction methodology discussed in \S~\ref{sec:intro} and provides an interpretation of concurrent 
behaviour in choreographies.
% , and an interpretation 
% of concurrent behaviour in choreographies.
% % 
% % 
% % CC is an attempt at defining a is tailored to be a realistic base model for the paradigm, meaning that:
% % (i) it offers the correct-by-construction methodology discussed in \S~\ref{sec:intro};
% % (ii) it distinguishes choreographic programming as a standalone paradigm, by providing an interpretation of 
% % concurrent
% % behaviour for choreographic programs that does not depend on other models;
% % (iii) it is realistically implementable as a real-world programming language.
%
%\begin{enumerate}
%\item it offers the correct-by-construction methodology from \S~\ref{sec:intro};
%\item it distinguishes choreographic programming as a standalone paradigm, by providing an interpretation of concurrent behaviour for choreographic programs that does not depend on other models;
%\item it is realistically implementable as a real-world programming language.
%\end{enumerate}
The key first-class elements of CC are \emph{processes} and \emph{sessions},
respectively representing endpoints that execute concurrently and the conversations among them. 
The basic statement of choreographic programs, ranged over by $C$, is a communication:
\[
\com{\pid p.e}{\pid q.x}{k};C
\]
which reads ``process $\pid p$ sends the value of expression $e$ to process $\pid q$, which receives it on variable 
$x$, over session $k$; then, the system executes the continuation choreography $C$''.
We comment the model by giving the following toy example on a replicated journaling file system.
\begin{example}[Replicated Journaling File System, write operation]
We define a choreography, denoted $C_{\m{jfs}}$, in which a client $\pid c$ uses a session $k$ to send some data 
to be written in a journaling file system replicated on two storage nodes.
\begin{displaymath}
\label{ex:cjfs}
\begin{array}{l@{\qquad}l}
	C_{\m{jfs}} \quad \stackrel{\m{def}}{=} &
	\begin{array}{l@{\quad}l}
	1. & \com{\pid c.data}{\pid j_1.data_1}{k};\\
	2. & \com{\pid c.data}{\pid j_2.data_2}{k};\\
	3. & \com{\pid j_1.\m{blocks}(data_1)}{\pid s_1.blocks_1}{k'};\\
	4. & \com{\pid j_2.\m{blocks}(data_2)}{\pid s_2.blocks_2}{k'};\\
	5. & \com{\pid j_1}{\pid c}{k};\\
	6. & \com{\pid j_2}{\pid c}{k}
	\end{array}
\end{array}
\end{displaymath}
In the choreography $C_{\m{jfs}}$, the client $\pid c$ uses session $k$ to send the $data$ to be written to two 
processes, $\pid j_1$ and $\pid j_2$, which we assume log the operation in their respective journals upon reception 
(Lines 1--2). The two journal processes then use another session, $k'$, to forward the data to be written to their 
respective processes handling the actual data storage, $\pid s_1$ and $\pid s_2$ (Lines 3--4). Finally, at the same 
time, processes $\pid j_1$ and $\pid j_2$ send an empty message on session $k$ to the client, in order to inform it 
that the operation has been completed (Lines 5--6).
\end{example}

\paragraph{Concurrency.}
Process identifiers ($\pid c$, $\pid j_1$, $\pid j_2$, $\pid s_1$ and $\pid s_2$ in our example) are key to formalising
concurrent behaviour in CC.
Observe Lines 3--4: since processes run in parallel, the communication 
between $\pid j_2$ and $\pid s_2$ in Line 4 could be completed before the communication between $\pid j_1$ and $\pid 
s_1$ in Line 3.
In CC, the semantics of the sequential operator is thus relaxed by a
syntactic \emph{swapping congruence relation} $\swapC$, which allows two statements to be swapped if they do not 
share any processes. For example, the choreography $C_{\m{jfs}}$ 
would be equivalent to a choreography $C'_{\m{jfs}}$, denoted $C_{\m{jfs}} \swapC C'_{\m{jfs}}$, where 
in $C'_{\m{jfs}}$ Lines 3 and 4 are exchanged.
In~\cite{M13:phd}, the relation $\swapC$ is validated
by showing that it corresponds to the typical interleaving semantics of the 
parallel operator found in process calculi.

\paragraph{Sessions and Typing.}
The communications in Lines 1--2, 5--6 and the communications in Lines 3--4 are included in different sessions, respectively $k$ and $k'$.
Each session represents a logically-separate conversation, as in other session-based calculi 
(e.g.,~\cite{HVK98,BCDLDY08}), and is strongly typed in CC with a typing
discipline that checks for adherence to 
protocols expressed as multiparty session types~\cite{HYC08}. We give an example of how protocols are mapped to 
choreographies in \S~\ref{sec:implementation}.
%	
%	 concrete processes in the 
%	choreography are matched to abstract roles in the protocol.
%	For example, by assigning the processes $\pid c$, $\pid j_1$ and $\pid j_2$ to the respective roles $\role C$, $\role 
%	{J1}$ and $\role{J2}$ we can type session $k$ in $C_{\m{jfs}}$ with the following protocol $G_k$:
%	\begin{displaymath}
%	\begin{array}{l@{\ }l}
%		G_k \ \stackrel{\m{def}}{=} &
%		\begin{array}{l@{\ }l}
%		\com{\role C}{\role{J1}}{\carr{\textbf{bytes}}};\ 
%		\com{\role C}{\role{J2}}{\carr{\textbf{bytes}}};\ 
%		\com{\role {J1}}{\role C}{\carr{\textbf{unit}}};\ 
%		\com{\role {J2}}{\role C}{\carr{\textbf{unit}}}
%		\end{array}
%	\end{array}
%	\end{displaymath}
%	Protocol $G_k$ types Lines 1--2 and 5--6 of the choreography $C_{\m{jfs}}$, matching the behaviour of the processes 
%	corresponding to the roles in the protocol. We assumed that the data communicated in Lines 1--2 is of type 
%	$\textbf{bytes}$. Empty messages are instead typed by $\textbf{unit}$.

\paragraph{Endpoint Projection.}
CC comes with an EPP that compiles choreographies to distributed implementations in terms of the 
$\pi$-calculus~\cite{CM13}.
The generated code follows that of 
the originating choreography, according to a small-step operational semantics.
As a corollary, the produced code is also deadlock-free: senders and receivers are always ready to 
communicate when they have to, as I/O actions cannot be mismatched in choreographies.
% 
% this holds because a choreography can never 
% ``get stuck'' as it is simply a sequential composition of statements where communications are expressed as atomic 
% statements that ensure that both the sender and receiver are ready to communicate when they have to. Thus, the typical 
% mismatches of I/O actions needed to create deadlocks cannot be written.

\paragraph{Modularity.}
% In CC and previous choreography models, programs are monolithic: they 
% describe the behaviour of all peers in the system and cannot reuse nor provide external
% libraries/services~\cite{BGGLZ06,QZCY07,LGMZ08,CHY12}.
% 
In~\cite{MY13}, we extend CC to support the implementation and reuse of external libraries/services (modular 
development), using a notion of external participants in sessions.
For example, we can split the choreography $C_{\m{jfs}}$ in two modules, a client choreography $C_{\m{cli}}$ and a 
server choreography $C_{\m{srv}}$:
% Interactions with external participants are typed to respect shared protocol specifications, to preserve the correctness 
% properties ensured by choreographies.
\begin{displaymath}
\label{ex:cjfs}
\begin{array}{lllll}
	C_{\m{cli}}\  \stackrel{\m{def}}{=}\ &
	\begin{array}{l@{\ }l}
	1. & \com{\pid c.data}{\role{J1}}{k};\\
	2. & \com{\pid c.data}{\role{J2}}{k};\\
	3. & \com{\role{J1}}{\pid c}{k};\\
	4. & \com{\role{J1}}{\pid c}{k}
	\end{array}
	& \quad &
	C_{\m{srv}}\ \stackrel{\m{def}}{=}\  &
	\begin{array}{l@{\ }l}
	1. & \com{\role C}{\pid j_1.data_1}{k};\\
	2. & \com{\role C}{\pid j_2.data_2}{k};\\
	3. & \com{\pid j_1.\m{blocks}(data_1)}{\pid s_1.blocks_1}{k'};\\
	4. & \com{\pid j_2.\m{blocks}(data_2)}{\pid s_2.blocks_2}{k'};\\
	5. & \com{\pid j_1}{\role C}{k};\\
	6. & \com{\pid j_2}{\role C}{k}
	\end{array}
\end{array}
\end{displaymath}
The choreographies above refer to each other using references to external processes, e.g., $\role{J1}$ in 
$C_{\m{cli}}$ is a reference to process $\pid j_1$ in $C_{\m{srv}}$. Separate choreography modules can be compiled and 
deployed separately, with the guarantee that their generated implementations will interact with 
each other as expected.
% Choreographies are 
% Typing sessions with behavioural types, such as global protocols, supports a particular way of 

\paragraph{Extraction.}
In~\cite{CMS14}, we present a proofs-as-programs Curry-Howard correspondence between Internal Compositional 
Choreographies (ICC, a simplification of CC) and a generalisation of Linear Logic~\cite{G87}, inspired by~\cite{CP10}.
ICC is a first step in defining a canonical model for choreographies and formalising logical reasoning on 
choreographic programs.
In such correspondence, EPP is formalised as a transformation between logically-equivalent proofs, one corresponding to 
a choreography program and the other corresponding to a $\pi$-calculus term. The transformation is invertible, yielding 
a procedure for automatically extracting the choreography that a $\pi$-calculus term typed with linear logic is 
following.

%% file: implementation.tex
% The language model formalised in CC is not sufficient in itself of choreographic programming, as 
The Choreography Calculus (CC), along with related work on models for choreography 
languages~\cite{QZCY07,LGMZ08,CHY12}, offers insight on how choreographic programming can be formally understood as a 
self-standing paradigm.
To practically evaluate choreographic programming, we developed the Chor programming 
language\footnote{\url{http://www.chor-lang.org/}}, an open source prototype implementation of CC~\cite{M13:phd}.
% The Chor language is an open source prototype implementation of CC, developed 
% for the real-world evaluation of choreographic programming~\cite{M13:phd}.
% 
% it is an abstract model and cannot be used for the real-world evaluation of 
% choreographic programming.
% In this section, we discuss the Chor programming language~\cite{M13:phd}, an open 
% source\footnote{\url{http://www.chor-lang.org/}} prototype implementation of CC.
% 

\begin{figure}[t]
\centering
\includegraphics[scale=0.56]{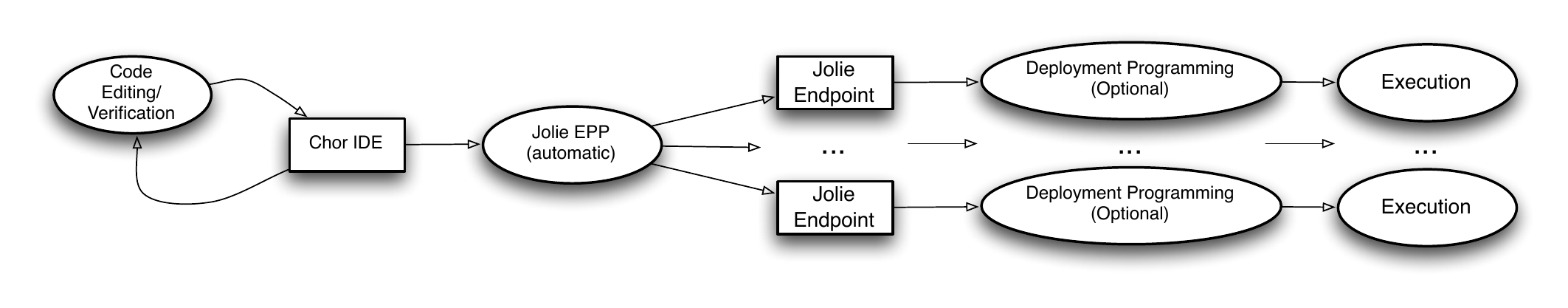}
\caption{Chor, development methodology (from~\cite{CM13}).}
\label{fig:chor_methodology}
\end{figure}
\begin{figure}[t]
\centering
\includegraphics[scale=0.4]{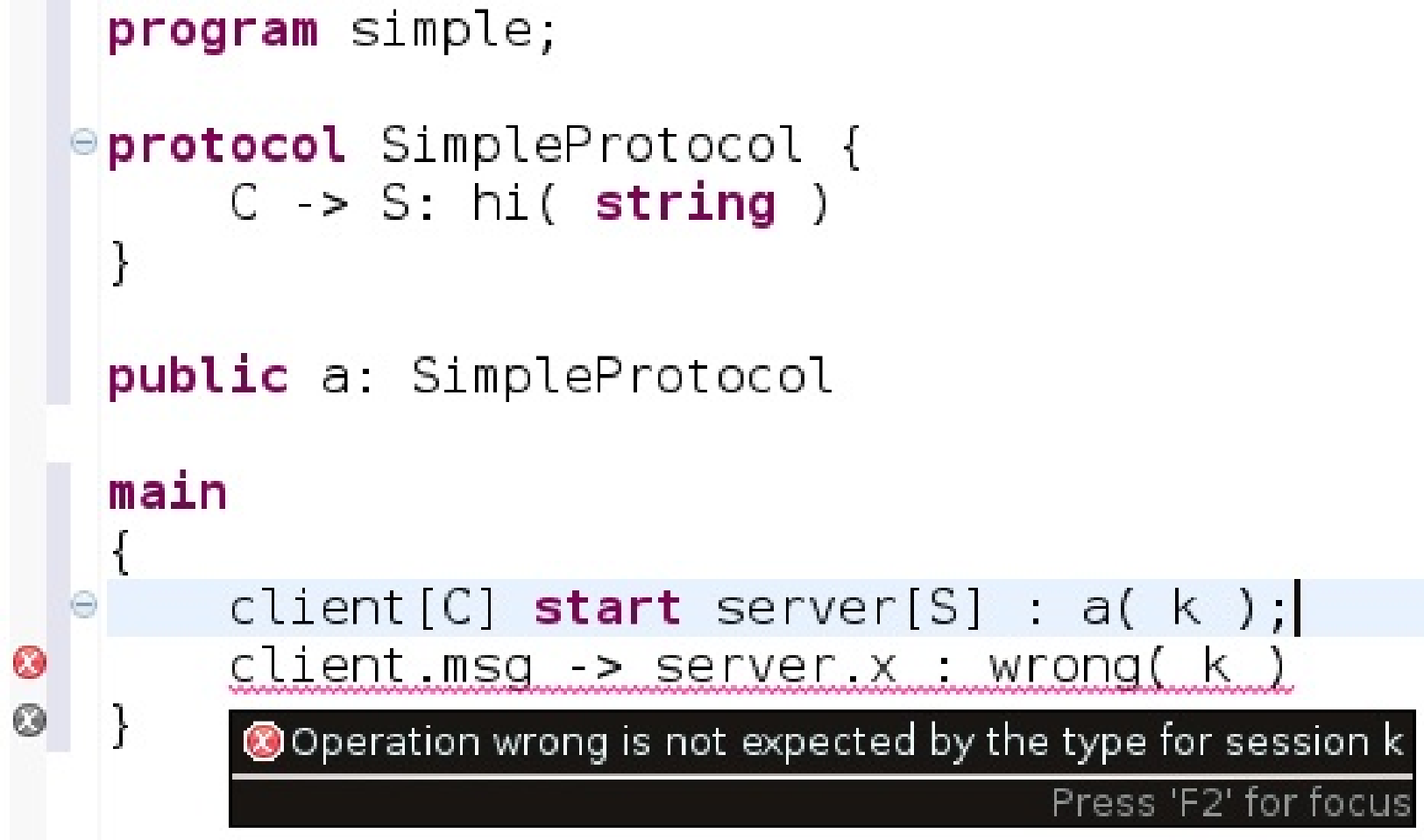}
\caption{Chor, example of error reporting (from~\cite{M13:phd}).}
\label{fig:chor_error}
\end{figure}
In Chor, the correct-by-construction methodology of choreographic programming is proposed as a concrete software 
development process, depicted in Fig.~\ref{fig:chor_methodology}.
Choreographies are written using an Integrated Development Environment (IDE), which visualises on-the-fly
errors regarding syntax and protocol verification, as in the screenshot in Fig.~\ref{fig:chor_error}.
Then, a choreography can be projected to executable code via an implementation of EPP that follows the 
ideas of CC. In this case, the target language is Jolie\footnote{\url{http://www.jolie-lang.org/}}~\cite{MGZ14}.
Once the compiler has generated the Jolie programs for the endpoints described in the choreography, the developer can 
customise their deployments. This is done using the Jolie primitives
% After the code for each Jolie endpoint is generated by the compiler, the developer can optionally customise their 
% deployments by using the different integration capabitilies offered by Jolie,
for integrating with standard communication protocols and technologies, which do not alter the behaviour of the 
code generated by the Chor compiler. The resulting code can finally be executed using the 
Jolie interpreter.

In Chor, the syntax from CC is extended with operation names for communications (as in Web 
Services~\cite{bpel}) and data manipulation primitives. As an example, we show an extended implementation of 
the scenario from Example~\ref{ex:cjfs}.
\setChor
%\begin{lstlisting}
%protocol ClientWrite {
%	C -> J1: write( string );
%	C -> J2: write( string );
%	J1 -> C: ok( void );
%	J2 -> C: ok( void )
%}
%
%protocol StoreWrite {
%	J1 -> S1: write( string );
%	J2 -> S2: write( string )
%}
%
%define write
%	( c, j1, j2, s1, s2 )
%	(
%	  k[ ClientWrite : c[C], j1[J1], j2[J2] ],
%	  k2[ StoreWrite : j1[J1], j2[J2], s1[S1], s2[S2] ]
%	)
%{
%	ask@c( "Data?", data );
%	c.data -> j1.data : write( k );
%	c.data -> j2.data : write( k );
%	
%	j1.data -> s1.data : write( k2 );
%	j2.data -> s2.data : write( k2 );
%	
%	j1 -> c : ok( k );
%	j2 -> c : ok( k )
%}
%\end{lstlisting}
%\todo{ comment ... }
% 
\begin{lstlisting}
protocol Write {
	C -> J1: { write(string);
									C -> J2: write(string);
									J1 -> C: ok(void);
									J2 -> C: ok(void),
             writeAsync(string);
									C -> J2: writeAsync(string)
	}
}

protocol Store { J1 -> S1: write(string);
                 J2 -> S2: write(string)  }

define computeBlocks(j1, j2) { /* ... */ }

define write(c, j1, j2, s1, s2)
(k[ Write:c[C], j1[J1], j2[J2] ],
 k2[ Store:j1[J1], j2[J2], s1[S1], s2[S2] ]) {
	if (sync)@c {
		c.data -> j1.data : write(k);
		c.data -> j2.data : write(k);
		computeBlocks( j1, j2 );	
		j1.blocks -> s1.blocks : write( k2 );
		j2.blocks -> s2.blocks : write( k2 );
		j1 -> c : ok( k );
		j2 -> c : ok( k )
	} else {
		c.data -> j1.data : writeAsync( k );
		c.data -> j2.data : writeAsync( k );
		computeBlocks( j1, j2 );	
		j1.data -> s1.data : write( k2 );
		j2.data -> s2.data : write( k2 )
	}
}
\end{lstlisting}
We briefly comment the program above, referring the reader to~\cite{M13:phd} for a more complete description of Chor.
Procedure \lstcode{write} implements the behaviour of the processes from Example~\ref{ex:cjfs}.
The sessions \lstcode{k} and \lstcode{k2} ($k$ and $k'$ in Example~\ref{ex:cjfs}) are typed by the protocols 
\lstcode{Write} and \lstcode{Store} respectively.
In Line 19, the client \lstcode{c} checks its internal variable \lstcode{sync} to determine whether the write 
operation should be synchronous or not. In the first case we proceed as in Example~\ref{ex:cjfs}. Otherwise, process 
\lstcode{c} uses a different operation \lstcode{writeAsync} to notify the others that it does not expect a confirmation 
message at the end.

%% file: related.tex
The idea of using choreography-like descriptions for communication behaviour
has been used for a long time, for 
example in software engineering~\cite{msc}, security~\cite{CVB06,BN07,BCDFL09}, and 
specification languages for business processes~\cite{wscdl,BPMN}.

The development of the formal models that we described in \S~\ref{sec:models}
was made possible by many other previous 
works on languages for expressing communication behaviour.
The notion of session in CC is a variation of that presented in~\cite{BCDLDY08}
for a process calculus.
The theory of modular choreographies was inspired by the article~\cite{BCVV12},
where types for I/O actions are mixed 
with types for global communications, and by Multiparty Session Types~\cite{HYC08}, from which we took the type language 
to interface 
compatible choreographies. Interestingly, combining multiparty session types with
choreographies yields a type inference 
technique and a deadlock-freedom analysis that do not require additional machinery as in other works in the context of 
processes~\cite{BCDLDY08}.
The criteria for a correct Endpoint Projection (EPP) procedure was investigated in many 
settings, e.g., in~\cite{QZCY07,BZ07,LGMZ08,CHY12}.

The Chor language and its compiler have already been used as basis for implementing
other projects. For example, 
AIOCJ~\cite{PGLMG14} is a choreographic language supporting the update of executable code at runtime, equipped with a 
formal calculus that ensure deadlock-freedom~\cite{PGGLM14}.
Choreographies have also been applied for the design of communication protocols.
In particular,
Scribble is a specification language for protocols written from a global viewpoint~\cite{YHNN13}, which can be used to 
generate correct-by-construction runtime monitors (see, 
e.g.,~\cite{NYH13}).

%% file: conclusions.tex
%	
%	In their own investigation on choreographic descriptions, Qiu et al. write that ``it seems that a lot of work should be done 
%	before a widely accepted result in this field''~\cite{QZCY07}; Aalst et al. 
%	also report that a ``real'' choreography language still has to be 
%	proposed~\cite{ADHRVW05}.
We presented some recent efforts aimed at kickstarting the development of choreographic programming as a fully-fledged 
programming paradigm. While the 
paradigm holds potential, there is still a lot of work to be done before reaching a productive real-world programming 
framework. We describe below some possible research directions, some of which are planned
for in the current research project behind Chor, the CRC project\footnote{\url{http://www.chor-lang.org/}}.

\paragraph{Integration.} A key factor for the
adoption of choreographic programming will be interoperability with 
existing software. Chor can be extended with local computation primitives that would interact with 
libraries written in other programming languages, e.g., Java or Scala, similarly to how it is done in 
Jolie~\cite{MGZ14}.

\paragraph{Classification.} Just like there are
many different language models for different aspects of concurrent 
programming, e.g., code mobility and multicast, it should be possible to similarly extend choreography models. 
This suggests a potential benefit in having systematic classifications of choreography languages, to observe the 
effect that such extensions have on expressiveness and see how far the correct-by-construction methodology can be 
applied.
% For example, one such classification could be based on computability (see, e.g.,~\cite{BGZ09}).
% 
% \paragraph{Expressiveness.}
% The model in CC could be extended with features
% that make writing distributed algorithms more natural, such as general 
% recursion. Choreographies are potentially a candidate for writing concurrent algorithms and check that they are 
% deadlock-free (e.g., distributed sorting).

\paragraph{Exceptions.} Introducing exception
handling in choreography program raises the issue of coordinating many 
participants in a global escape (as in~\cite{CHY09}), and whether a suitable strategy can always be found, statically or 
at runtime.

\paragraph{Formal Implementation.}
% The EPP procedure in Chor uses data (protocol headers) instead of abstract channelsto route communications, as in many 
% other enterprise frameworks~\cite{bpel}, instead of channels
% 
The EPP procedure in CC is based on $\pi$-calculus
channels, but its implementation in Chor uses data (protocol headers) to route messages
instead, as in many other enterprise frameworks~\cite{bpel}. To the best of the author's knowledge, realising
$\pi$-calculus channels using data-based message routing has still to be formally investigated, and the implementation 
of Chor could provide an initial stepping stone in such a study.